\documentclass[iop]{emulateapj}
\usepackage{comment}
\usepackage{color}
\usepackage{xspace}
\usepackage{ulem}

\newcommand{\lya}        {Ly$\alpha$\xspace}
\newcommand{\civ}        {\ion{C}{4}\xspace}
\newcommand{\heii}       {\ion{He}{2}\xspace}

\newcommand{\ha}         {H$\alpha$\xspace}

\newcommand{\oiii}       {[\ion{O}{3}]\xspace}
\newcommand{\unitcgssb}  {erg\,s$^{-1}$\,cm$^{-2}$\,arcsec$^{-2}$\xspace}

\newcommand{\unitcgslum} {erg\,s$^{-1}$\xspace}
\newcommand{\unitlpco}   {K\,km\,s$^{-1}$\,pc$^2$\xspace}

\newcommand{\kms}        {\,km\,s$^{-1}$\xspace}

\newcommand\E[1]         {$\times$10$^{#1}$}

\newcommand\bb      {\phn}
\newcommand\co[2]   {CO {\sl J}\,=\,#1--#2}
\newcommand\ra[3]   {#1$^{\rm h}$\,#2$^{\rm m}$\,#3$^{\rm s}$}
\newcommand\dec[3]  {#1\degr\,#2\arcmin\,#3\arcsec}

\newcommand\ICO[2]  {$I_{\rm CO(#1-#2)}$\xspace}
\newcommand\LpCO[2] {${L^{\prime}}_{\rm\!\!CO(#1-#2)}$\xspace}

\newcommand\cii     {[\ion{C}{2}]\xspace}
\newcommand\um      {\ifmmode\mu{\rm m}\else$\mu${\rm m}\fi\xspace}
\newcommand\sig     {$\sigma$\xspace}

\newcommand\lpco    {${L^{\prime}}_{\rm\!\!CO}$\xspace}
\newcommand\lco     {$L_{\rm CO}$\xspace}

\newcommand\lfir    {$L_{\rm FIR}$\xspace}

\newcommand\msun    {$M_{\sun}$\xspace}
\newcommand\lsun    {$L_{\sun}$\xspace}
\newcommand\mhtwo   {$M({\rm H_2})$\xspace}
\newcommand\htwo    {H$_{\rm2}$\xspace}

\newcommand\perbeam {beam$^{-1}$\xspace}

\newcommand\dvlya   {\ifmmode\Delta v_{\rm Ly\alpha}\else$\Delta v_{\rm Ly\alpha}$\xspace\fi}

\slugcomment{Accepted for publication in ApJ.}
\shorttitle{Molecular Gas in a L\lowercase{y}$\alpha$ Blob}
\shortauthors{Yang et al.}

\begin{document}

\title{
Pinpointing the Molecular Gas within a L\lowercase{y$\alpha$} Blob at \lowercase{$z$} $\sim$ 2.7 \\ 
}

\author{
   Yujin Yang\altaffilmark{1,2},
   Fabian Walter\altaffilmark{2},
   Roberto Decarli\altaffilmark{2},
   Frank Bertoldi\altaffilmark{1},
   Axel Weiss\altaffilmark{3}, \\
   Arjun Dey\altaffilmark{4,5},
   Moire K. M. Prescott\altaffilmark{6},
   Toma B{\u a}descu\altaffilmark{1} 
}

\altaffiltext{1}{Argelander Institut f\"ur Astronomie, Universit\"at Bonn, Auf dem H\"ugel 71, 53121 Bonn, Germany}
\altaffiltext{2}{Max-Planck-Institut f\"ur Astronomie, K\"onigstuhl 17, Heidelberg, Germany}
\altaffiltext{3}{Max-Planck-Insitut f\"ur Radioastronomie, Auf dem H\"ugel 69, D-53121 Bonn, Germany}
\altaffiltext{4}{National Optical Astronomy Observatory, 950 N. Cherry Ave., Tucson, AZ 85719}
\altaffiltext{5}{Radcliffe Fellow, Radcliffe Institute of Advanced Study, Harvard University 10 Garden Street, Byerly Hall Cambridge, MA 02138}
\altaffiltext{6}{Dark Cosmology Centre, Niels Bohr Institute, University of Copenhagen, Juliane Maries Vej 30, 2100 Copenhagen \O, Denmark}


\begin{abstract}

We present IRAM Plateau de Bure Interferometer observations of the
CO(3--2) and CO(5--4) line transitions from a \lya blob at $z$ $\sim$
2.7 in order to investigate the gas kinematics, determine the location
of the dominant energy source, and study the physical conditions of the
molecular gas.
CO line and dust continuum emission are detected at the location of a
strong MIPS source that is offset by $\sim$1.5\arcsec\ from the \lya peak.
Neither of these emission components is resolved with the 1.7\arcsec\
beam, showing that the gas and dust are confined to within $\sim$7\,kpc
from this galaxy.  No millimeter source is found at the location of the
\lya peak, ruling out a central compact source of star formation as the
power source for the \lya emission.
Combined with a spatially-resolved spectrum of \lya and \heii, we
constrain the kinematics of the extended gas using the CO emission
as a tracer of the systemic redshift.  Near the MIPS source, the
\lya profile is symmetric and its line center agrees with that of CO
line, implying that there are no significant bulk flows and that the
photo-ionization from the MIPS source might be the dominant source of
the \lya emission.  In the region near the \lya peak, the gas is slowly
receding ($\sim$100\kms) with respect to the MIPS source, thus making
the hyper-/superwind hypothesis unlikely.
We find a sub-thermal line ratio between two CO transitions,
\ICO{5}{4}/\ICO{3}{2} = 0.97 $\pm$ 0.21. This line ratio is lower than
the average values found in high-$z$ SMGs and QSOs, but consistent
with the value found in the Galactic center, suggesting that there is
a large reservoir of low-density molecular gas that is spread over the
MIPS source and its vicinity.

\end{abstract}


\section{Introduction}
\setcounter{footnote}{0}

Giant Ly$\alpha$ nebulae (also known as ``\lya blobs'') are large
(50--100~kpc) spatially extended regions emitting copious amounts
of Ly$\alpha$ emission [$L$(\lya)$\sim$ 10$^{43-44}$ \unitcgslum]
\cite[e.g.,][]{Keel99, Steidel00, Francis01, Matsuda04, Matsuda11, Dey05,
Saito06, Smith&Jarvis07, Ouchi09, Prescott09, Prescott12a, Yang09, Yang10,
 Erb11}.  They may represent sites of massive galaxy formation and their
early interaction with the intergalactic medium. However, the questions
of what powers these gigantic gas halos and whether the surrounding
gas is outflowing from or infalling into the embedded galaxies
are still debated. The proposed scenarios include photo-ionization
by AGNs \citep{Geach09}, shock-heated gas by galactic superwinds
\citep{Taniguchi&Shioya00}, cooling radiation from cold-mode accretion
\citep{Fardal01, Haiman00, Dijkstra&Loeb09, Goerdt10}, and resonant
scattering of \lya from star-forming galaxies \citep{Steidel11, Hayes11}.
However, observations of \lya blobs currently allow no firm conclusions
about their nature. For example, less than 20\% of blobs are found to
contain X-ray luminous AGN \citep{Geach09} implying that a powerful
AGN is not always required for producing the extended halo \cite[see
also][]{Yang09}.  Yet \citet{Overzier13} argue strongly that most are
powered by AGN.  It has long been believed that \lya blobs undergo intense
dusty starbursts like submillimeter galaxies \citep{Ivison98, Chapman04,
Geach05}, but recent studies show that most \lya blobs are not as
luminous at rest-frame far-infrared (FIR) wavelengths as submillimeter
galaxies \citep{Yang12, Tamura13}.

The molecular gas content in \lya blobs has to date remained
unconstrained, despite numerous attempts to detect CO emission
\citep{Chapman04, Yang12, Wagg12a}.  Observations of the molecular gas
component in these systems are key to determining the physical conditions
in the star-forming gas, the star formation efficiency, and constraining
various relevant timescales and the role of feedback.
CO emission lines are also a useful tool as a tracer of the systemic
redshift, without which it is difficult to constrain the gas
kinematics within a given system.  For example, previous studies of
the same \lya blob using only \lya emission line have led to radically
different conclusions about the gas kinematics, suggesting that gas is
outflowing in stellar or AGN winds \citep{Wilman05}, inflowing due to
gas accretion \citep{Dijkstra06b}, or even static \citep{Verhamme08}.
These discrepancies mainly arise because the systemic velocities of the
galaxies within the \lya blob are not directly measured, but rather need
to be assumed \cite[see also][]{Bower04,Weijmans09}.
While other nebular lines in the UV/optical such as \oiii and \ha have
been used to constrain the systemic velocity within a small sample
of \lya blobs \cite[e.g.,][]{Yang11,McLinden13}, they are strongly
susceptible to extinction and are only detectable from the ground over
a restricted redshift range due to atmosphere.  In such cases, the far
infrared emission lines such as CO and \cii 158\um are ideal probes to
determine the systemic redshift of the system, should they harbor a cold
molecular component. 


In this paper, we discuss new Plateau de Bure Interferometer
(PdBI) observations of the CO $J$ = 3\,$\rightarrow$\,2 and $J$ =
5\,$\rightarrow$\,4 line transitions from one of the most luminous \lya
blobs \cite[LABd05;][]{Dey05}.
Originally discovered due to its strong {\it Spitzer} MIPS 24\um
flux, LABd05 has been the subject of many multi-wavelength studies.
{\it Spitzer} and high resolution {\sl HST} optical/NIR observations
show that this giant \lya nebula is composed of a strong MIPS source
and $\sim$17 small compact galaxies \citep{Prescott12b}.  The peak of
the \lya emission is substantially offset from all the galaxies (e.g.,
1.5\arcsec\ from the MIPS source), but is coincident with a detection
in \heii emission.  \citet{Prescott08} found that LABd05 inhabits an
overdense environment, suggesting that it may be the progenitor of a
rich galaxy group or low-mass galaxy cluster.
The SED of the bright MIPS source within LABd05 can be explained by an
AGN-dominated template with a FIR luminosity of \lfir(40--1000\um) =
4\E{12}\lsun \citep{Dey05, Bussmann09, Yang12}, although it is not yet
clear whether the far-IR luminosity is driven primarily by star-formation
or AGN activity \citep{Colbert11}.
The MIPS source contributes the bulk of the bolometric luminosity from
the region and has a very extreme rest-frame UV-to-mid-IR color that
characterizes it as a heavily dust-obscured galaxy \cite[DOG;][]{Dey08,
Prescott12b}.
On the other hand, the kinematics of \lya-emitting gas have not been
fully constrained.  While the Keck Low Resolution Imaging Spectrometer
(LRIS) longslit spectroscopy shows that there is a monotonic (and
approximately linear) velocity gradient across the \lya blob (maybe due
to rotation, outflow or infall), the interpretation of this signature was
ambiguous because of the lack of the systemic redshift of the embedded
galaxies, possibly the center of the gravitational potential of the blob
\citep{Dey05}. Previous attempts to detect the CO emission in LABd05 using
the IRAM-30m yielded only upper limits \citet{Yang12}.  Here, we present
new PdBI interferometric observations of LABd05 that reveal detections
of the dust continuum emission and CO emission lines from the system.

This paper is organized as follows.  In \S\ref{sec:observation}, we
describe our PdBI observations.  In \S\ref{sec:detection}, we present
the detection of molecular gas from LABd05.  In \S\ref{sec:voffset}, we
constrain the gas kinematics of the \lya-emitting gas using the systemic
redshift derived from CO and \heii lines.  In \S\ref{sec:COSED}, we use
the CO line SED to study physical conditions in star-forming regions.
In \S\ref{sec:SED_LABd05}, we present the FIR SED and derive constraints
on the dust properties of LABd05.  In \S\ref{sec:model}, we discuss
some plausible physical models for LABd05.  Section \ref{sec:conclusion}
summarizes our conclusions. Throughout this paper, we adopt the following
cosmological parameters: $H_0$ = 70\,${\rm km\,s^{-1}\ Mpc^{-1}}$,
$\Omega_{\rm M}=0.3$, and $\Omega_{\Lambda}=0.7$.

\section{Observations and Data Analysis}
\label{sec:observation}


PdBI observations of the \co{3}{2} transition for LABd05 were carried out
using the WideX receiver (3\,mm band) in June -- August 2012 (project ID:
{\tt W04E}) over 9 observing sessions. At the same time, we targeted
another \lya blob SSA22--LAB18 (R.A. = \ra{22}{17}{28.99}, decl. =
\dec{+00}{07}{51.2}), a source originally discovered by \citet{Matsuda04}
and known to host an SMG with $S_{850\um}$\,=\,11\,mJy \citep{Geach05}.
These two \lya blobs were selected mainly because they are the brightest
in the (sub)mm among the sources accessible from PdBI.

Data were collected in the compact D configuration with four or five
antennae and baselines ranging between 24\,m and 112\,m. The total
bandwidth of the dual polarization mode was 3.6\,GHz corresponding
to $\sim$12000\,\kms. We tuned the receiver to $\nu_{\rm obs}$ =
94.58\,GHz and 84.55\,GHz, the redshifted CO(3--2) frequency at $z$
= 2.656 (LABd05) and $z$ $\approx$ 3.090 (SSA22-LAB18), respectively.
The on-source integration times equivalent to the full six-antenna array
were 4.7 and 5.5 hours for LABd05 and SSA22--LAB18, respectively. The
water vapor ranged between 5\,mm and 12\,mm on different sessions.
%
%
The data were calibrated through observations of bandpass (3C279,
2013+370, 2200+420) phase/amplitude (1308+326, 1417+385, 1504+377)
and flux calibrators (MWC349) for LABd05.
For SSA22--LAB18, we used bandpass (3C454.3, 2013+370, 2200+420),
phase/amplitude (2230+114, 2145+067) and flux (MWC349, 2200+420)
calibrators.
Phase calibrators were observed every $\sim$22 minutes. The primary
beam sizes are 53\arcsec\ and 59\arcsec\ with the synthesized beams of
6\farcs4$\times$4\farcs2 (P.A.=54\degr) and 6\farcs9$\times$5\farcs2
(PA=29\degr) for LABd05 and SSA22-LAB18, respectively.


The CO(3--2) line emission is detected with high significance
($\sim$6\sig) from LABd05 while the observations of SSA22-LAB18 resulted
only in a 3$\sigma$ upper limit of $\approx$0.27 Jy\,\kms (assuming a
line-width of 400\,\kms). We obtained additional PdBI WideX observations
of the \co{5}{4} transition from LABd05 on 2012 December 22 and 30
(project ID: {\tt W0AC}).
Data were collected in C configuration with six antennae and baselines
ranging between 24\,m and 176\,m.  The receiver was tuned to $\nu_{\rm
obs}$ = 157.617\,GHz (2\,mm band) and the total bandwidth was 3.6\,GHz, 
corresponding to $\sim$6800\,\kms.  The on-source observing time
was 5.2 hours. The water vapor ranged between 2\,mm and 10\,mm.
The primary beam and the synthesized beam sizes were 32\arcsec\ and
1\farcs8\,$\times$\,1\farcs6, respectively.  For calibration, we used
bandpass (3C273, 3C279) phase/amplitude (1424+366, 1417+273), and flux
(3C273, 3C279) calibrators.


\begin{figure*}
\epsscale{1.0}
\plotone{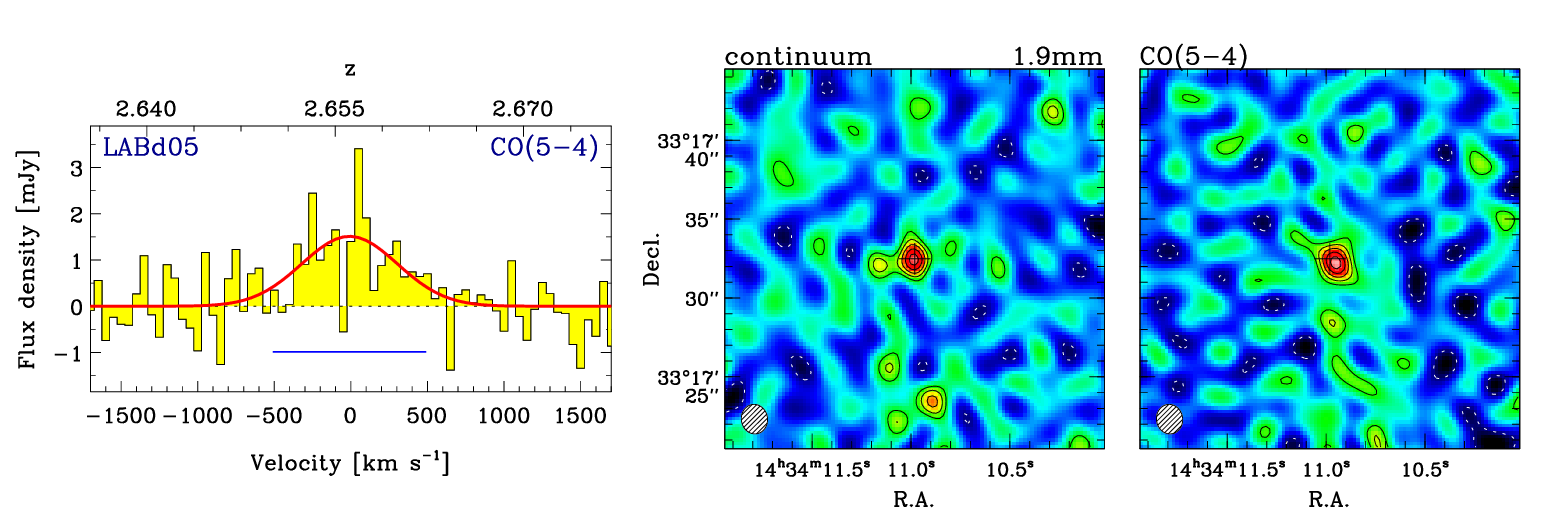}
\plotone{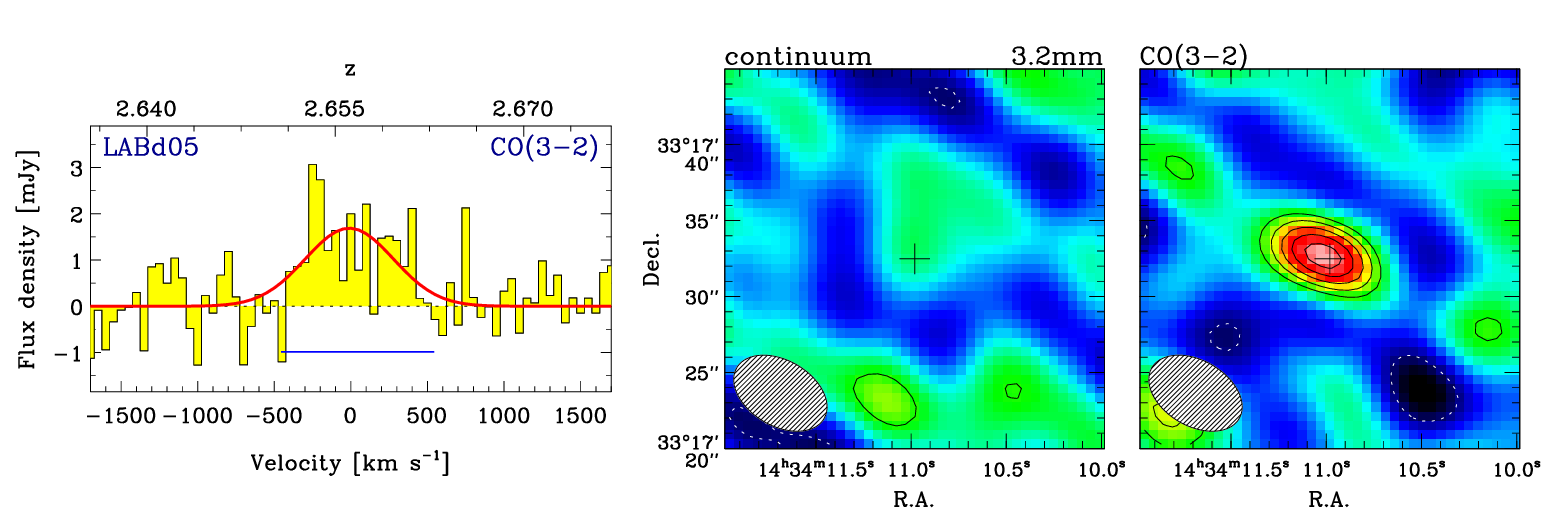}
\caption{
PdBI CO(5--4) and CO(3--2) line observations of LABd05.  Both the
dust continuum and CO line images are unresolved down to 1.7\arcsec\
($\sim$13.5\,kpc).
({\it Left})
Continuum-subtracted CO spectra with the velocity resolution of
50\,\kms. The solid (red) curves represent the Gaussian fits to the line
profiles. The (blue) horizontal bars represent the velocity intervals
used to create the line-only images.
({\it Middle})
Cleaned continuum images generated from the line-free spectral regions.
({\it Right})
Cleaned CO line images after subtraction of the continuum
emission.  The crosses indicate the phase centers (galaxy
\#36; R.A.~= {14$^{\rm h}$\,34$^{\rm m}$\,10\fs981}, decl.~=
{$+$33\degr\,17\arcmin\,32\farcs48}).  The contours are spaced for 2,
3, 4, 5, 6\,\sig ({\it solid}\,) and $-$2 and $-$4\,\sig ({\it dotted
line}\,).  The sizes of the synthesized beams are shown in the bottom
left corner of each panel. The field of view (FOV) of each image panel
is $\sim$25\arcsec.
}
\label{fig:CO}
\end{figure*}



The data were reduced with {\tt CLIC} and {\tt MAPPING} within
the {\tt GILDAS} software package.{\kern-0.1em}\footnote{\tt
http://www.iram.fr/IRAMFR/GILDAS} The data processing program used
water vapor monitoring receivers at 22\,GHz on each antenna to
correct the measured amplitudes and phases for short-term changes in
atmospheric water vapor.  The typical rms errors in phase calibration
are 20\degr--45\degr\ and 10\degr--30\degr\ for the 2\,mm and 3\,mm band
observations, respectively.  Typical uncertainties in the flux scales and
overall calibration are about 10\%.  We weighted visibilities according
to the inverse square of the system temperature, and applied natural
weights when creating maps.  We achieved an rms sensitivity of 0.4 --
0.5\,mJy\,\perbeam per 100\,\kms channel for both targets and transitions.

\section{Results}
\label{sec:result}

\subsection{Detection of CO in a \lya Blob}
\label{sec:detection}

We detect the CO line emission from the molecular gas associated with
LABd05.
In Figure \ref{fig:CO}, we show the CO spectra, the cleaned dust continuum
and the integrated CO line images for LABd05.  Since neither dust
continuum nor CO line emission is detected in SSA22-LAB18, we present
the results for SSA22-LAB18 in the Appendix.  We extract the spectra
from the brightest pixel in the image cubes because the source is not
resolved in the current data.  For the CO(5--4) line observations where
the underlying dust continuum is detected, we subtract the continuum in
the $uv$ space before extraction.  We fit single Gaussian profiles to the
spectra to measure the line fluxes and widths.  We produce the CO line
images by integrating the channels indicated with horizontal lines in
the left panels of Figure \ref{fig:CO}, which are $\sim$1.3$\times$FWHM
of the lines. The CO(3--2) and CO(5--4) lines are detected at 6\sig and
7\sig level for this velocity interval, respectively. The properties of
the CO lines and dust continuum are listed in Table \ref{tab:CO}.


\begin{figure*}
\centering
\includegraphics[width=0.45\textwidth]{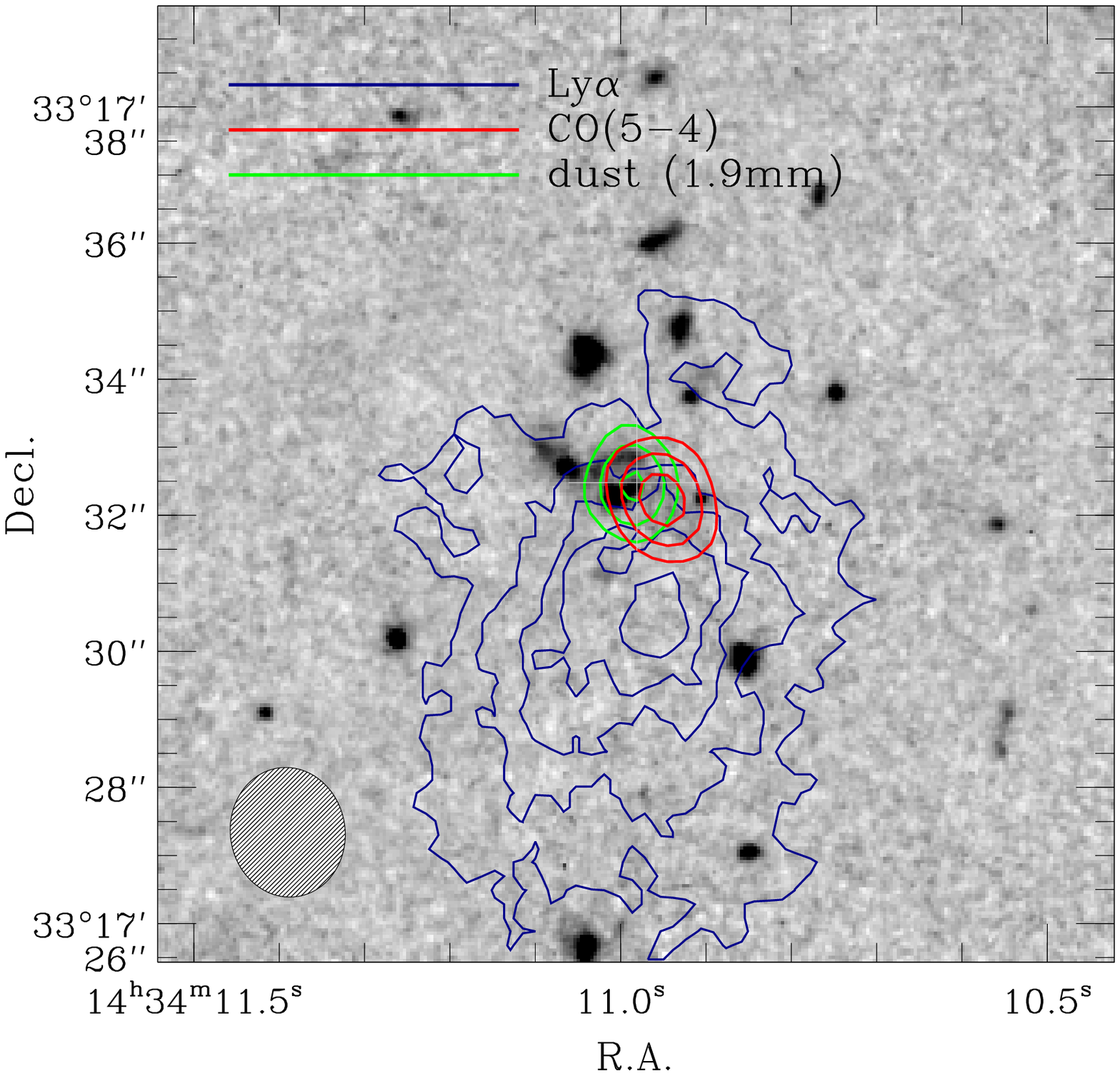} 
\includegraphics[width=0.45\textwidth]{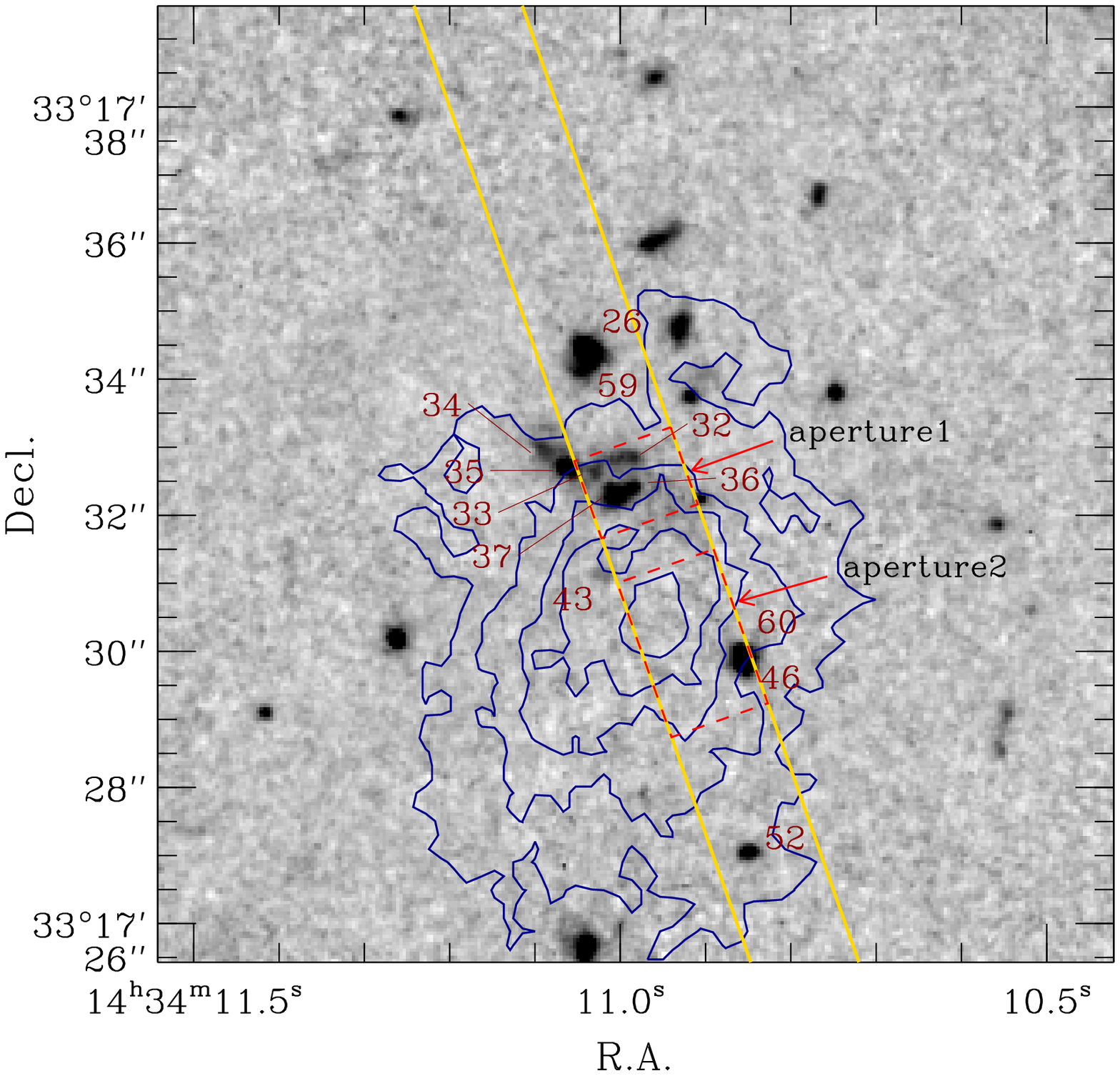} 
\caption{
({\it Left})
{\sl HST} {\sl VJH} composite image of LABd05 \cite[adopted
from][Fig.~2]{Prescott12b} overlayed with \lya, 1.9\,mm dust continuum,
and CO(5--4) contours. The FOV is 14\arcsec\ ($\approx$110\,kpc).
The cross indicates the phase center of the PdBI observation,
which is the position of the galaxy \#36.  Green, red, blue lines
represents 1.9mm dust continuum, CO(5--4) and \lya lines, respectively.
For clarity, we show only 3, 4, 5, 6, 7\sig contours for the dust
continuum and CO emission. \lya contours are plotted for 1, 3, 5, 7,
9\,$\times$\,$10^{-17}$\unitcgssb.
Our PdBI submm observations pin down the exact location of the most
energetic source within the \lya blob The dust and molecular gas is
concentrated on one galaxy (\#36) rather than spread out over the
\lya blob. In particular, there is no source found at the peak of \lya
emission.
({\it Right})
{\sl HST} {\sl VJH} composite image overlayed with the position
of 1.5\arcsec-wide Keck LRIS slit \cite[PA=19.4\degr;][]{Dey05}.
The galaxies that fall within the slit are labeled in red. The two dashed
boxes are the extraction apertures 1 and 2, centered on the MIPS source
(\#36) and the \heii-emitting region, respectively. The galaxies \#46
and \#60 are not members of LABd05, but background galaxies.
}
\label{fig:hst}
\end{figure*}


The PdBI high-resolution continuum observations pinpoint the
location of the most energetic source within LABd05.  In Figure
\ref{fig:hst}, we show the 1.9mm continuum and CO(5--4) line contours
on an {\sl HST} {\sl VJH} composite image.  The dust continuum
and CO lines are detected at the phase center, the position of the
galaxy \#36 (R.A.~= {14$^{\rm h}$\,34$^{\rm m}$\,10\fs981}, decl.~=
{$+$33\degr\,17\arcmin\,32\farcs48}), which was identified from the
deep optical and near-infrared {\sl HST} observations as the most likely
counterpart to the strong MIPS source \citep{Prescott12b}.
The dust continuum and CO line images are unresolved down to our
synthesized beam (1.7\arcsec; 13.5\,kpc) and are coming from galaxy \#36,
or at least from the $r$ $\approx$ 7\,kpc region around this position.
The center of the CO emission is slightly offset from the phase center
by 0\farcs4\,$\pm$\,0\farcs3, but this offset is not statistically
significant.\footnote{The uncertainty on the position of the CO
emission is derived from the {\tt UVFIT} routine in {\tt GILDAS}.}

No millimeter source is detected at the location of the peak of the
Ly$\alpha$ emission, which is offset by $\sim$1.5\arcsec\ (12\,kpc in
projected distance) from the MIPS source.  We place a 3\sig upper limit
of $S_{\rm 1.9mm}$ $<$ 0.15 mJy\,\perbeam corresponding to \lfir $<$
1.6\E{12}\lsun or SFR $<$ 260\,\msun{yr$^{-1}$}, assuming the dust
properties presented in Section \ref{sec:SED_LABd05}. It is therefore
unlikely that the nebula is powered by a luminous, dust-obscured
galaxy located at the position of the \lya centroid.

The CO line spectra show that the MIPS source (galaxy \#36) is indeed
located at the same redshift as the extended \lya emission ($z$ $\simeq$
2.66).  Note that prior to this study, the redshift of this MIPS source
had only been estimated from a weak detection of PAH emission in low
S/N {\it Spitzer} InfraRed Spectrograph (IRS) spectra \citep{Colbert11}.
The redshifts and widths of the CO(5--4) and CO(3--2) lines agree to
within the uncertainties. By averaging the two measurements, we determine
the systemic redshift to be $z$ = 2.6560 $\pm$ 0.0005, corresponding to a
velocity uncertainty of $\sim$40\kms. We also fitted the two CO spectra
simultaneously by forcing both profiles to have the same redshift and
line-width, but different peak intensities, and found that the resulting
redshift and uncertainties were consistent with the above. Therefore,
we are confident with our redshift determination.  We will discuss the
\lya, \civ, \heii and CO line profiles in detail in the next section.

Given the FIR luminosity of the source, the measured CO line-width
(FWHM $\approx$ 700\,\kms) is relatively large, suggesting that the
embedded galaxy (\#36) is quite massive.  \citet{Carilli&Walter13}
compiled the \lfir and the FWHM of CO lines for high-$z$ SMGs and
QSOs (their figure~5).  While there is a large scatter and no clear
correlation between \lfir and FWHM, the LABd05 is located in the upper
part of the distribution of FWHMs for the sources with $\log$\,\lfir
$\sim$ 12.5, suggesting that LABd05 contains a galaxy whose dynamical
mass is comparable to that of QSOs and SMGs.  The line-width of LABd05
is larger than that of more quiescent normal star-forming galaxies or
color-selected star-forming galaxies.
A model-dependent mass estimate for the galaxy \#36 is $M_{\rm dyn}$
$\equiv$ $5\sigma^2 R / G$ = 3--7$\times$$10^{11}$\msun, where the range
reflects the range of size estimates for the CO emitting region: the
optical size of the galaxy \#36 \cite[$R_e$ = 0.39\arcsec;][]{Prescott12b}
and the beam size of CO(5--4) observations as an upper limit ($R$
= 7\,kpc).
Higher resolution CO observations could yield a more accurate mass
estimate by spatially resolving the velocity structure of the CO-emitting
region and by determining the inclination of the galaxy.

\subsection{Gas Kinematics from \lya--CO Velocity Offset}
\label{sec:voffset}


\begin{figure*}
\epsscale{1.00}
\epsscale{1.05}
\plotone{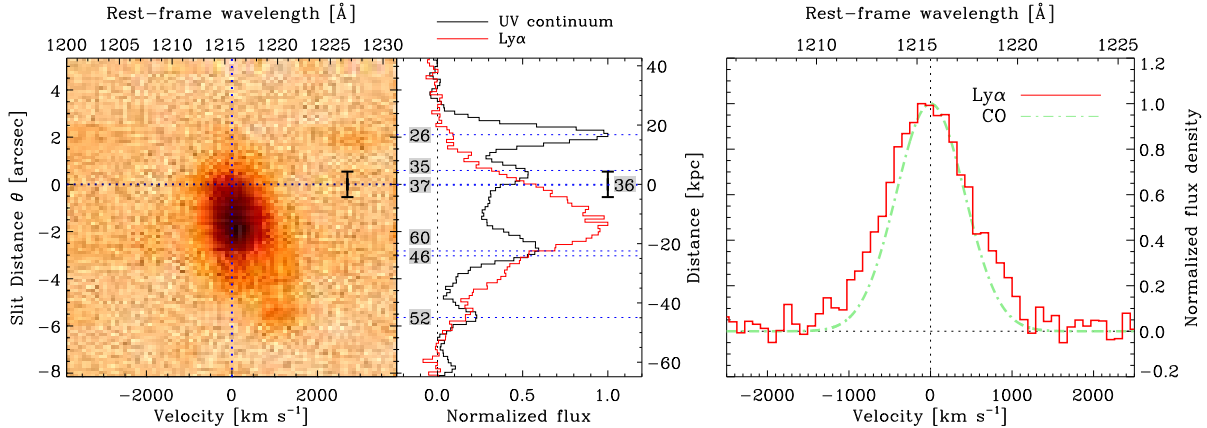}
\caption{
({\it Left})
Keck LRIS 2--D \lya spectra and the spatial profile of the UV continuum
along the slit.  The vertical line represents the systemic redshift of
the CO-detected source (galaxy \#36). The galaxies that fall within
the LRIS slit are identified on the right. The black and red histograms
indicate the spatial profiles of the rest-frame UV continuum and the
\lya, respectively.
({\it Right})
1--D \lya spectrum (red histogram) extracted from the region indicated
by the vertical black bar on the left panel (aperture 1), i.e. the
approximate location of galaxy \#36.  The green dot-dashed line is the
Gaussian fit to the CO profile that was convolved to the spectral
resolution of the optical spectra.  While the \lya profile is broader
(FWHM = 1040\,\kms) than the CO line (FWHM = 700\,\kms), the \lya
profile is symmetric and the line center agrees very well with that of
the CO line.  The velocity offset between \lya and CO is \dvlya $\approx$
$-$25$\pm$15\kms, much smaller than the extreme values ($\sim$1000\kms)
predicted by hyper-/superwind model for \lya blobs.  We conclude that
there is no significant bulk motion around the MIPS source (\#36).
}
\label{fig:spec2d}
\end{figure*}


We next compare the CO and Ly$\alpha$ line profiles to investigate
the kinematics of the extended \lya emission.  Previous studies of
LBGs and compact \lya emitters have used velocity offsets between the
optically thick \lya and the optically thin nebular emission lines (e.g.,
\oiii) as diagnostics for for the kinematics of the surrounding medium
\citep{Steidel04, Steidel10, McLinden11, Hashimoto13}. We refer the
readers to \citet{Yang11} for the detailed descriptions and caveats of
this technique in the context of \lya blobs.  In short, in an optically
thick medium with no bulk flows, a double-peaked \lya profile with similar
intensity emerges because of the resonant scattering. If the medium is
outflowing, the blue peak of the \lya profile is depressed or diminished;
thus the brighter peak of the \lya profile appears redshifted relative
to the systemic velocity. In the case of inflowing gas, the brighter \lya
peak appears blueshifted instead \cite[e.g.,][]{Dijkstra06a,Verhamme06}.

In this study, we use CO emission lines instead of nebular lines to
determine the systemic velocity, because the CO, as a tracer of the
cold molecular gas component, should be the best tracer of the systemic
redshift of a galaxy \cite[e.g.,][]{Greve05}. Here we assume that the
observed CO emission is originating from the interstellar medium within
galaxy \#36. As will be discussed in \S\ref{sec:COSED}, LABd05 might have
a large reservoir of diffuse molecular gas, possibly due to outflows or
galaxy interactions, which may affect the centroid of CO line.  However,
we find that the line profiles of the two CO transitions --- which may
trace different phases of the molecular gas --- show almost identical
line centers and no evidence for complex kinematics, suggesting that
our assumption is valid.

To compare velocity centers between the CO lines and the rest-frame UV
emission lines, we re-analyze the Keck LRIS longslit spectra that were
first presented in \citet{Dey05}.  This spectrum was obtained with a
1.5\arcsec-wide slit with PA = 19.4\degr\ centered on the MIPS source.
In the following, we compare the CO, \lya, \civ and \heii line profiles
at two apertures along the slit: one line of sight (LOS) directly toward
the galaxy \#36 (aperture 1 in Fig.\,\ref{fig:hst}), and a sight line
toward the \heii-emitting gas near the \lya peak (aperture 2).  For a
proper comparison between radio and optical spectra, all the spectra
are placed in a vacuum wavelength and the local standard of rest (LSR)
velocity frame.\footnote{The difference between LSR and heliocentric frame
is negligible in our application, but we adopt the LSR frame because most
extragalactic radio/submm observations adopt the LSR frame as a default.}


Figure \ref{fig:hst} (right) shows the location of the longslit, the
extraction apertures, and the sources identified in {\sl HST} images
\citep{Dey05, Prescott12b}. We label the sources which fall within
the slit under $\sim$1\arcsec\ seeing condition (adopting the naming
convention of \citealt{Prescott12b}).  To determine the positions of the
two apertures along the slit, we collapse the spectrum over a wavelength
range of [1240\AA\ -- 1720\AA].  Figure \ref{fig:spec2d} (left) shows
the 2--D spectrum and the spatial profile of rest-frame UV continuum.
We bootstrap the location of the MIPS source (\#36) using the several
sources (\#26, \#35, \#46, \#52, \#60) that are bright enough to be
detected in the continuum profile ($m_{\rm F606W}$ $\lesssim$ 26.5).
The blended galaxies (\#26, \#59) and (\#46, \#60) correspond to the
galaxy A and B in \citet{Dey05}, respectively.  While the membership
of galaxy A was spectroscopically confirmed before, galaxy B is not a
member, but a background source at $z$ $\sim$ 3.27.

Aperture 1 is centered on the MIPS source (\#36). \citet{Prescott12b}
show that this source has a strong Balmer/4000\AA\ break, and a very
concentrated light profile, and that it is surrounded by numerous
neighbouring galaxies and diffuse emission indicating that these galaxies
might undergo merging events.
Aperture 2 is centered on the \heii-emitting region, $\approx$2.5\arcsec\
away from galaxy \#36 along the slit, and also includes the \lya peak
which is located $\approx$1.6\arcsec\ south of the MIPS source (\#36).
For this part of \lya blob, no galaxy is identified in the entire
wavelength range from the rest-frame UV to FIR, thus the \lya and \heii
emission originate from the extended gas itself. This \heii region
is extended at least over 0\farcs6. We refer readers to \citet{Dey05}
and \citet{Prescott12b} for further details.

%
%

\subsubsection{Kinematics around the MIPS Source}

We compare the CO and \lya line profiles extracted at the position of
galaxy \#36 (aperture 1). Figure \ref{fig:spec2d} shows the \lya profile
from a 1.2\arcsec\ (9 pixels) aperture centered on the galaxy \#36.
For comparison, we overlay the Gaussian fit to the CO profile after
convolving it with the spectral resolution of LRIS optical spectra
($\sigma_{\rm instr}$ = 295\,\kms).  While the \lya profile is broader
(FWHM = 1040\,\kms) than the CO line (FWHM = 700\,\kms), the profile
is symmetric and the line center agrees with that of the CO line within
the uncertainty (\dvlya = $-25$ $\pm$ $15$\,\kms).


\begin{figure*}
\epsscale{1.00}
\epsscale{1.05}
\plotone{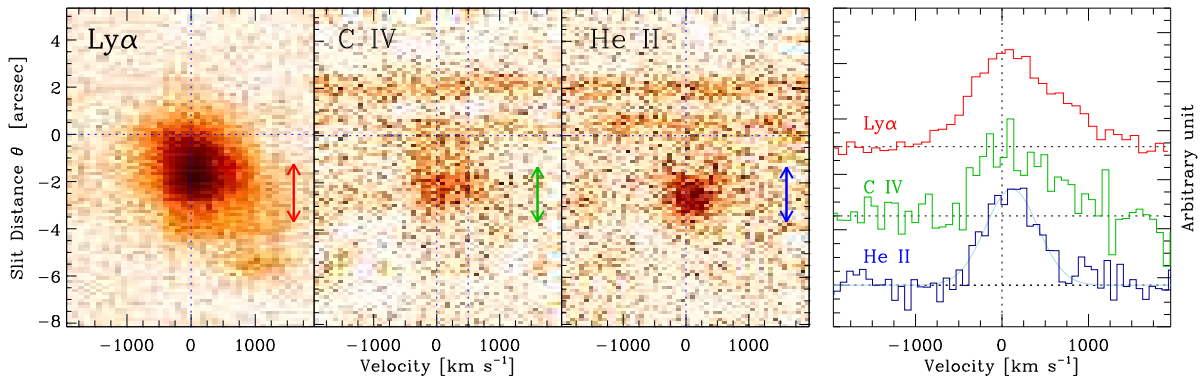}
\caption{
({\it Left})
2--D \lya, \civ, and \heii spectra. As in Fig.~\ref{fig:spec2d}, vertical
and horizontal lines in each panel represent the location and systemic
velocity of the MIPS source (\#36).
({\it Right})
1--D spectra extracted from the \heii-emitting region (aperture 2;
$\theta$ = $-$2.5\arcsec), which roughly coincides with the peak of \lya
surface brightness ($\theta$ = $-1.6$\arcsec). The spectra are shifted
along the $y$--axis by an arbitrary unit for clarity.  In contrast to
aperture 1 (Fig.~\ref{fig:spec2d}), the \lya profile at this location is
broader and more asymmetric with a red wing.  The profile of the blended
\civ doublet appears to be similar to the \lya profile.  The optically
thin \heii line is redshifted by a small amount ($\sim$110\,\kms) against
the CO line center of aperture 1, indicating that the \heii-emitting
region is receding relative to the MIPS source.
}
\label{fig:spec2d_HeII}
\end{figure*}


A symmetric single-peaked \lya profile located exactly at the systemic
velocity is not often observed in \lya-emitting galaxies, which typically
show redshifted asymmetric profiles or double-peaked profiles (e.g.,
\citealt{Steidel10, Kulas12, Yamada12}, but see \citealt{Yang11}, for
exceptions).  Given the observed single \lya peak, the possibilities
are (1) the \lya profile is intrinsically doubled-peaked with similar
intensity but two peaks are blended because of low spectral resolution or
(2) the \lya photons produced in the MIPS source (either due to star
formation or AGN) are {\it not} seriously affected by the resonant
scattering in the medium between the galaxy and the observer.
We find that the first scenario is unlikely because another Keck LRIS
spectrum taken with a higher resolution ($\sigma_{\rm instr}$ = 197\,\kms)
also shows a symmetric \lya profile. This spectrum was also obtained by
\citet{Dey05} with PA = 0\degr, but we do not present it here because
the slit does not cover the \heii-emitting region.
Though unlikely, this first scenario would imply that the gas around the
MIPS source is roughly static or that there is no significant bulk motion.
The second scenario implies that there is less neutral \ion{H}{1} gas at
the systemic velocity of the MIPS source, possibly due to photo-ionization
by the ionizing source.  If this is the case, the velocity shift of the
\lya line would reflect gas motions.  In either case, we conclude that
the \lya profile along the LOS toward galaxy \#36 is consistent with
roughly static gas without any significant bulk flows.
However, we note that we cannot rule out the possibility of gas motions
tangential to our LOS, thus hiding any strong outflow signatures.

\subsubsection{Kinematics of \heii-emitting Gas}

We compare the \lya, \heii, \civ profiles with the CO profile for the
\heii region (aperture 2) to investigate how this \heii-emitting gas
is moving relative to the MIPS source. For this part of \lya blob,
the extended optically thin \heii emission must originate from the
extended gas itself.  For optically thick \lya lines, \lya photons
could be produced in situ at the same region as \heii, or they could be
produced by other galaxies or within other parts of the gaseous halo,
and subsequently resonantly scattered into this region.

Figure \ref{fig:spec2d_HeII} shows the 2--D spectra of \lya, \civ and
\heii and the 1--D spectra extracted from the \heii-emitting region
(aperture 2 centered at $\theta$ = $-$2.5\arcsec, where $\theta$ is the
projected distance from the MIPS source along the slit). This extraction
aperture is indicated as a vertical arrow in the left panel.
At this aperture, the peak of the \lya profile still agrees with the
systemic velocity of the MIPS source ($\Delta v$ = 45\,$\pm$\,26 \kms),
but the \lya profile is asymmetric with a red wing that extends up
to 1500\kms.  We find that the velocity half-width at half-maximum
(HWHM  = 1350\kms) on the red side of the profile is roughly twice
broader than the blue side (HWHM = 600\kms).
The profile of the blended \civ doublet lines appears to be similar
to the \lya profile, but the low S/N does not allow us to carry out
a detailed analysis.
On the other hand, the optically thin \heii line is redshifted by a small
amount ($\Delta v$ = 110\,$\pm$\,25\kms) against the systemic velocity
of the \#36, indicating that the \heii-emitting gas is moving relative
to the MIPS source.
In general, it would not be possible to determine whether the
\heii region is approaching or receding relative to the MIPS source
because we do not know whether this gas is located in front of or
behind the galaxy.  Luckily, the \lya profile from the same region has
an asymmetric red wing, indicating that there must be receding material
by which the \lya photons are back-scattered.  Therefore, we conclude
that it is most likely that the \heii-emitting gas near the \lya-peak is
receding in the local velocity frame and located behind the MIPS source.

The symmetric \lya profile at the location of the MIPS source (i.e.,
in aperture 1) and the relatively small LOS velocity offset of the \heii
emitting region ($\sim$ 100\,\kms) together suggest that LABd05 does not
have extreme outflows as predicted by the hyper-/superwind hypothesis
\citep{Taniguchi&Shioya00} and that shock-heating does not play a major
role in powering \lya emission in this source.  This extreme outflow
model, which requires an outflow speed up to $\sim$1000\,\kms, has been
proposed to explain the large extent of \lya blobs.
Of course, we cannot completely rule out the possibility of strong
outflows oriented tangential to the line-of-sight, but we note that
\citet{Dey05} showed that the \civ/\heii line ratio is inconsistent with
shock ionization. A large statistical study of \lya blob gas kinematics
is required to better constrain the gas geometry.
We note that no evidence for extreme outflows was found among a total
of eight \lya blobs with a wide range of \lya luminosity (Y.\,Yang
in preparation).
We will further discuss the physical nature of LABd05 in detail in
Section \ref{sec:model}.

\subsection{CO SED of LABd05}
\label{sec:COSED}

\begin{figure}
\epsscale{1.00}
\epsscale{1.20}
\plotone{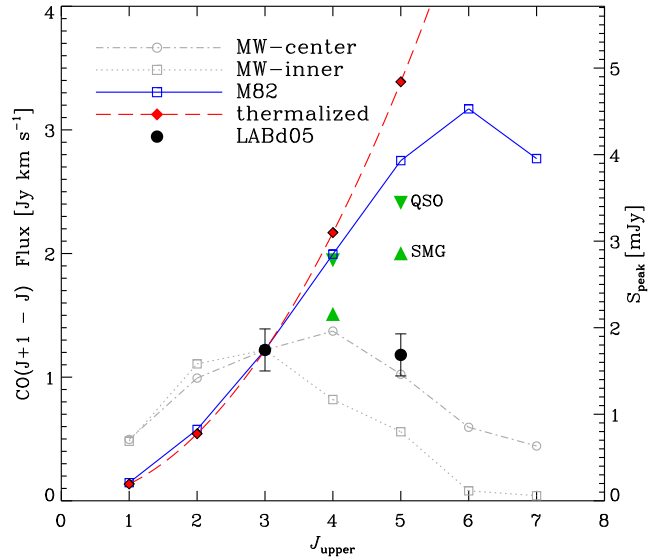}
\caption{
CO line SED of LABd05. The CO line flux is plotted versus the rotational
quantum number $J_{\rm upper}$ of the CO molecule. The left $y$--axis
shows the measured line fluxes, the right $y$--axis the peak flux density
of the lines in LABd05.  For comparison, we show the CO SEDs for Milky
Way (dot-dashed line), a starburst galaxy (M82; solid line) and a
fully thermalized CO gas ($I_{\rm CO}$ $\propto$ $J_{\rm upper}^2$;
dashed line) which are normalized at $I_{\rm CO(3-2)}$ of LABd05.
The triangles represent the average values of high-$z$ SMGs and QSOs. The
CO(5--4)/CO(3--2) line ratio of LABd05 is lower than those of other
high-$z$ sources with similar \lfir.
}
\label{fig:COSED}
\end{figure}

In combination with radiative transfer models such as the large velocity
gradient (LVG) approximation or the photo-dissociation region (PDR)
model, multiple CO transitions (CO SED) can be used to constrain the
physical conditions of the molecular gas in the star-forming region
\cite[e.g.,][]{Weiss07}.  As a first step to studying the properties
of star-forming gas in the galaxy that dominates the bolometric
luminosity of the \lya blob, we show the CO SED of LABd05 in Figure
\ref{fig:COSED}.  For comparison, we show the CO SEDs for the Milky
Way \citep{Fixsen99}, the center of a local starburst galaxy M82
\citep{Weiss05}, and an optically thick thermalized gas [\ICO{J+1}{J}
$\propto$ $(J+1)^2$].  The average values for high-$z$ QSOs and SMGs
compiled by \citet{Carilli&Walter13} are also shown for $J_{\rm upper}$
= 4 and 5.  All CO SEDs are normalized at the integrated line flux
\ICO{3}{2} = 1.22 Jy\,\kms of LABd05.

The CO SED of LABd05 is unusual given its high FIR luminosity
(\lfir $\sim$ 4\E{12}\lsun). We find a sub-thermal line ratio
between two transitions, \ICO{5}{4}/\ICO{3}{2} = 0.97 $\pm$ 0.21
[\LpCO{5}{4}/\LpCO{3}{2} = 0.35 $\pm$ 0.08] which is smaller than the
ratios for the center of M82 \cite[2.24;][]{Weiss05} and for fully
thermalized gas (2.78). This ratio is even lower than the average
values of high-$z$ QSOs (1.97) and SMGs (1.64). The compilation of
\citet{Carilli&Walter13} includes a total of 9 SMGs and QSOs that
have both CO(3--2) and CO(5--4) line flux measurements with a range of
\ICO{5}{4}/\ICO{3}{2} = 1.4 -- 2.4, thus LABd05's line ratio is lower
than those of SMGs and QSOs.\footnotemark\ The observed line ratio is
higher than the value (0.50$\,\pm\,$0.1) for the inner part of the MW
($2.5\degr < |\,l\,| < 32.5\degr$), but consistent with the value found
toward the Galactic center \cite[0.84$\,\pm\,$0.06;][]{Fixsen99}.

\footnotetext{These measurements are compiled from \citet{Barvainis97,
Frayer99, Neri03, Weiss03, Weiss09, Hainline04, Tacconi06, Alloin07,
Swinbank10, Riechers11a, Danielson11, Schumacher12}.}


A caveat to our measurement of the line ratio is that the beam sizes for
the two transitions are different: $\approx$5\arcsec\ and 1.7\arcsec\
for CO(3--2) and CO(5--4), respectively.  Therefore, it is conceivable
that our CO(5--4) line observations might have out-resolved any extended
CO emission, leading to an underestimate of the total CO(5--4) flux
and a lower line ratio. We tested whether any extended emission can
be recovered from our C configuration data. Tapering visibilities with
$\sim$50m width boosted the CO(5--4) line flux by only $\sim$20\% with
a new synthesized beam ($\sim$4\arcsec): \ICO{5}{4} = 1.36$\,\pm\,$0.3,
but still consistent with the original measurement.  Note that in order
for LABd05's ratio to be consistent with that of SMGs, the CO(5--4)
flux has to be boosted by $\sim$70\%, i.e., more than $\sim$1/3 of the
total flux has to lie outside the 1\farcs7 beam.
However, since the gas traced by CO(5--4) is likely to be more spatially
compact, it is unlikely to be missed (or resolved-out) by the larger
baselines and the 1.7\arcsec\ beam used here.
Future observations of additional CO transitions (e.g., $J$ =
1$\rightarrow$0, 7$\rightarrow$6) and/or re-observation of CO(5--4) line
with a compact array configuration are required to further constrain
the CO SEDs of LABd05.

With these caveats in mind, we consider two possible scenarios for the
CO excitation in LABd05 and their implications.
First, the LABd05's CO emission could arise from a single gas phase,
i.e., the CO SED of LABd05 can be approximated with a LVG model with a
single gas density ($n$) and temperature ($T_{\rm kin}$).
In this case, one can show that the star-forming gas in
this Ly$\alpha$ blob has lower density and/or temperature than what has
been observed in QSOs and SMGs with similar \lfir.
Low CO excitation has been observed for high-$z$ galaxies, but only for
a handful of near-IR-selected galaxies (the so-called BzK galaxies)
at $z\sim1.5$ that have lower \lfir ($\sim$10$^{12}$\lsun) than SMGs
and LABd05 \cite[e.g.,][]{Dannerbauer09}.


Second, the CO SED of LABd05 could be composed of two components (a)
a dense molecular gas reservoir typical of high-$z$ QSOs and (b)
an additional diffuse gas component around this system.  There is
now increasing evidence that SMGs contain both low and high-excitation
components of molecular gas \cite[e.g.,][]{Weiss05, Ivison11, Danielson11,
Riechers11b}.
For example, \citet{Riechers11b} show that, in two SMGs (SMM J09431+4700
and SMM J13120+4242), a low-excitation component similar to the MW CO
SED is required to explain the excess of low-$J$ emission.  However,
we find that even in these studies, the low-excitation component is not
strong enough to yield a flat \ICO{5}{4}/\ICO{3}{2} line ratio.
Instead, we find that the flat (5--4)/(3--2) line ratio of LABd05 is
more similar to that of the local starburst galaxy M82. \citet{Weiss05}
show that while the central starburst disk of M82 follows an almost
thermalized CO SED up to $J$=4 or 5 (Fig.~\ref{fig:COSED}), the total
integrated CO SED is less excited due to low-$J$ transitions dominated by
the diffuse molecular gas located in the outflow and in tidal streamers.
Their best-fit LVG model shows that if M82 were observed unresolved
at high-$z$, it would have \ICO{5}{4}/\ICO{3}{2} = 0.83 similar to the
values observed in LABd05.

Without the complete CO SED, especially the $J$=1$\rightarrow$0
transition, it is not possible to distinguish these two scenarios
discussed above. However, in either case, we conclude that there must be a
significant reservoir of low density molecular gas associated with LABd05.

\subsection{Dust Properties of LABd05}
\label{sec:SED_LABd05}


\begin{figure}
\epsscale{1.00}
\epsscale{1.20}
\plotone{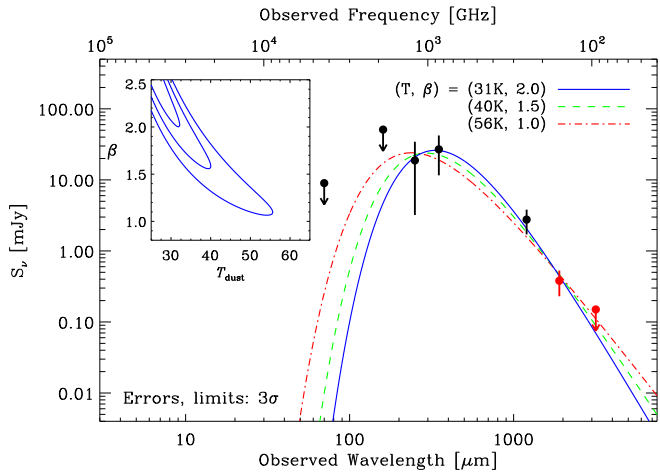}
\caption{
Spectral energy distribution of LABd05 from mid-IR to mm wavelengths.
The 1.9\,mm and 3.2\,mm data from this work (red) are combined with the
data from \citet{Yang12}.  The error bars and upper limits are given
at 3$\sigma$.  The left inset shows the likelihood distribution of $T_d$
and $\beta$. The contours represent the 1, 2 and 3\,$\sigma$ confidence
intervals.  We find $T_d$ = 54, 39, 31\,K for the fixed $\beta$ = 1,
1.5, and 2, respectively. While there is a degeneracy between $T_d$
and $\beta$, we find that models with colder dust temperature ($T_d$
$\lesssim$ 40\,K) are favored at 2\sig (96\%) confidence.
}
\label{fig:dust_SED}
\end{figure}


The dust properties of \lya blobs have been poorly constrained to date
because most \lya blobs have not been detected at mm and submm
wavelengths \citep{Matsuda07, Kohno08, Yang12, Tamura13}.
Using new continuum measurements, we update the far--infrared SED
of LABd05.  In Figure \ref{fig:dust_SED}, we show the 1.9\,mm flux
density and the upper limit on the 3.2\,mm continuum together
with the measurements presented in \citet{Yang12}.  We fit the
data with a modified blackbody SED with a dust temperature $T_d$
and an emissivity index $\beta$.  We find $T_d$ = 54, 39 and 31\,K
for the fixed emissivity index $\beta$ = 1, 1.5 and 2, respectively.
While the best-fit $T_d$ and $\beta$ are still degenerate, the colder
dust temperature ($T_d$ $<$ 41\,K) and large emissivity index ($\beta$
$>$ 1.5) are favoured at the 2\sig (96\%) confidence level. Our \lfir
estimate remains almost unchanged from \citet{Yang12}: \lfir(40--1000\um)
= (3.9$\,\pm\,$0.5)$\times$$10^{12}$\lsun. The updated dust mass is
$M_{\rm dust}$ = (1.5$\,\pm\,$0.6)\E{9}\msun. Note that the derived dust
mass is sensitive to the choice of $\beta$: if we adopt $\beta$ = 1.5,
$M_{\rm dust}$ will decrease by a factor of $\sim$4.


We find that the dust properties within the \lya blob are more consistent
with those of SMGs [($T_d$, $\beta$) = (35\,K, 1.6)] than with those of
QSOs \cite[47\,K, 1.6;][]{Beelen06}.
Recently, using the Wide-Field Infrared Survey Explorer (WISE) and a
MIR color selection technique, \citet{Bridge13} selected hyper-luminous
infra-red galaxies (\lfir $\gtrsim$ $10^{13}$ -- $10^{14}$\lsun), some of
which show extended \lya halos over 30--100\,kpc. These WISE-selected \lya
blobs have very warm dust temperatures, $T_d$ = 50--85\,K ($\beta = 1.5$),
much warmer than LABd05, typical SMGs and less luminous DOGs.
This warm temperature indicates that the dust is being illuminated by
an AGN, thus \citet{Bridge13} suggest that these WISE-selected HyLIRGs
are experiencing extremely powerful feedback from the AGNs and that
they are in a very brief evolutionary stage between dusty starbursts
and optical QSOs.
Given that the colder dust temperature in LABd05 is preferred by our SED
fit and that this dust temperature is consistent with those of DOGs and
SMGs rather than AGNs, we conclude that the dust heating in the FIR is
likely dominated by star formation in the MIPS source.

\subsection{Molecular Gas Mass in LABd05}
\label{sec:gas_mass}

The unusual CO excitation in LABd05 adversely affects our ability
to estimate the molecular gas mass. If we adopt the MW-like CO SED
and extrapolate CO fluxes to $J$ = 1, LABd05 will have \LpCO{1}{0} =
1.6\E{11} \unitlpco and \mhtwo = 6.4\E{11}\,($\alpha_{\rm CO}$/4.0)\,\msun
(Fig.~\ref{fig:COSED}). Here we adopt the nominal MW-like CO--to--\htwo
conversion factor.  On the other hand, if we assume that the
gas is still fully thermalized up to $J$ = 3, we obtain \mhtwo =
3.5\E{10}\,($\alpha_{\rm CO}$/0.8)\,\msun for \LpCO{1}{0} = 4.4\E{10}
\unitlpco assuming a ULIRG-like conversion factor.  The actual gas mass
\mhtwo and \LpCO{1}{0} values for LABd05 are expected to lie between
these two extreme values.  For these two \mhtwo estimates, we obtain
the gas-to-dust mass ratios, $M_{\rm gas}$/$M_{\rm dust}$ = 20 and 370,
which span almost the entire observed range of Local Group galaxies
\citep{Leroy11} and local ULIRGs \citep{Solomon97}.  Furthermore, the
\mhtwo estimate derived from the MW-like CO SED is comparable to or
even larger than the dynamical mass estimated from the CO line-width.
We conclude that putting constraints on the CO(1--0) line luminosity
of LABd05 and carrying out more detailed modeling of the CO SED will
be critical for constraining the properties of star formation in this
\lya blob.

\section{Discussion}

With new information on the gas kinematics in hand, we construct
physical models for LABd05 in this section. In the following, we begin
by summarizing the observational findings obtained from this work as
well as previous studies \citep{Dey05, Prescott12b}, which a successful
model must be able to explain:

\begin{enumerate}
\item
The \lya profile around the MIPS source (galaxy \#36) is symmetric and
peaks precisely at the systemic velocity of the galaxy as traced by
CO emission.

\item
The \heii-emitting gas (located $\sim$20\,kpc away from the MIPS source)
is receding with $\sim$100\,\kms with respect to the MIPS source.

\item
The \lya profiles become broader, more asymmetric, and more redshifted
with increasing distance from the MIPS source \citep{Dey05}.

\item
There are no galaxies detected in the UV-to-FIR at the peak of the
\lya emission, which is located at $\sim$12\,kpc from the MIPS source
\citep{Prescott12b}.

\end{enumerate}
%
The unknowns are (1) the mass distribution in LABd05 and (2) how
to interpret the relative motion between the MIPS source and the
\heii-emitting gas. Depending on the answers to these two unknowns,
we consider the following two models for LABd05.

\subsection{Physical Models for LABd05}
\label{sec:model}

The first model assumes that the mass (gas and dark matter) is centered
on the MIPS source and that the \heii/\lya-emitting gas is outflowing
from this galaxy.
In this picture, photo-ionization by an obscured AGN (the MIPS source)
is the dominant source for the extended \lya emission. 
The power-law MIR SED of the MIPS source and no obvious sign of AGN
activity in the optical spectrum (e.g., broad emission lines and high
ionization lines) indicates that the MIPS source might contain an
obscured AGN \citep{Dey05}.  Note that star formation is likely to
heat the dust within the MIPS source (confined within $\sim$7\,kpc),
but the radiation from the AGN provides energy for the \lya emission
over $\sim$100\,kpc scale.
As the MIPS source, plausibly the most massive member of LABd05, is also
located near the greatest concentration of galaxies, it likely lies at
or near the center of the gravitational potential.
A cone of radiation from the AGN photo-ionizes the surrounding gas toward
the south--west part of the \lya blob.  The presence of \civ and \heii
emission near the \lya peak suggests that this ionizing radiation should
be hard.
Since we do not observe any \lya or \heii emission towards
the northern side, in this model, the radiation from the AGN must be
anisotropic.
Note that this kind of anisotropic ionizing radiation is common among
high-redshift QSOs \cite[e.g.][]{Hennawi13, Farina13}.  Around the
AGN, the gas is highly ionized, leading to \lya being less affected by
resonant scattering off of neutral gas.  Therefore, the \lya profile
is single-peaked and symmetric for this LOS.  The surrounding gas is
outflowing with a relatively low speed ($\sim$100\kms in projection)
from the MIPS source due to either star-formation or AGN-driven winds.
Further from the AGN, the gas becomes more neutral, thus the \lya profiles
become broader and more asymmetric.  The observed monotonic velocity
shear across the \lya halo could be explained if the outflowing gas is
accelerating \cite[e.g.,][]{Steidel10} or radiative transfer effects,
as the \lya lines become more redshifted due to higher optical depths
at larger radii.  In this scenario, the diffuse molecular gas might
represent gas entrained in an outflow from the MIPS source.

The major challenge to this model is the fact that the \lya and
\heii emission peaks are not located at the location of the MIPS source
where illumination from the AGN and the gas density should be highest.
Given that the \lya surface brightness depends on complicated radiative
transfer effects, the anisotropic geometry of the radiation cone, and the
varying ionization state of the gas, it is not clear whether the observed
\lya surface brightness distribution can be reproduced by this model.
Detailed \lya radiative transfer calculations are required.


An alternative scenario is that the gas in LABd05 is distributed
symmetrically with a density profile that scales with the surface
brightness profile of \lya.  In other words, the potential is centered
around the location of the \lya peak and the galaxies are somehow
displaced from the center of the potential.  Because there are no
galaxies directly associated with this \lya peak, the baryonic component
is largely in the gas phase. The kinematics of the gas are reflected in
the observed \lya profiles.
The galaxies are forming from this halo gas, therefore they closely
follow the motions of the gas. In other words, the galaxies have only
small peculiar velocities relative to the overall gas kinematics. The
MIPS source is one of these galaxies forming within the nebula, thus
explaining the lack of any significant velocity offset between \lya
and CO.  It is undergoing a starburst and possibly AGN formation, and
is supplying ionizing photons to the gas.
In this case, the ionization from the MIPS source could be isotropic
and there could be also other ionizing sources. For example,
\citet{Prescott12b} found {\it diffuse} continuum emission near the
\heii region which, if powered by spatially extended star formation,
could suggest an additional source of ionization.  In this model,
the \lya emission is observed only on one side of the AGN because the
photo-ionization is bounded by the gas distribution.  The low density
molecular gas around the MIPS source may indicate that molecular gas is
forming within the MIPS source and fueling the star formation activity.
In other words, these observations show a massive galaxy in the process
of forming from the gas with rapid dust and molecular gas formation.

The major challenge in this model is how to explain the observed
\lya velocity structure, in particular, why the line-width of \lya
varies across the halo rather than being broadest at the center of the
nebula. Furthermore, it is not clear why the galaxies and gas in this
system are spatially offset.

With the limited kinematic data, it is difficult to discriminate between
these two models. Furthermore, it is still possible that there is an
obscured source at the location of the \lya peak just below the
detection limit of our mm observations (SFR $<$ 260\,\msun\,yr$^{-1}$;
3\sig). Alternately, a weak radio-jet from the MIPS source might be
responsible for the offset morphology between \lya and member galaxies,
similar to what has been observed in high-redshift radio galaxies.
To reveal the nature of LABd05, deeper spectroscopic observations with
high spectral resolution and deeper (sub)mm/radio continuum observations
will be required to further constrain the kinematics of the \lya- and
\heii-emitting gas and to search for possible energy sources.

\section{Summary and Conclusions}
\label{sec:conclusion}

We have obtained IRAM PdBI observations of the CO $J$ =
3\,$\rightarrow$\,2 and $J$ = 5\,$\rightarrow$\,4 line transitions from
a $z$=2.7 \lya blob (LABd05) in order to investigate the molecular gas
content and kinematics, to determine the location of the dominant energy
source, and to study the physical conditions of star-forming regions
within the \lya blob.

We detect CO line emission from the molecular gas associated with this
\lya blob. 
The CO line emission and the dust continuum are detected at the location
of a strong MIPS source, which is offset by $\sim$1.5\arcsec\ (12\,kpc
in projection) from the peak of \lya surface brightness distribution.
Neither the CO line nor the dust continuum emission is resolved with our
1.7\arcsec\ beam, showing that the molecular gas and dust are confined
to within a $\sim$7kpc region around the MIPS source. In addition, no
millimeter continuum source is found at the location of the \lya peak,
excluding the presence of a compact source of hidden star formation at
SFR $>$ 260\msun{yr$^{-1}$} which might be directly responsible for the
\lya emission.  The CO line spectra show that the MIPS source is indeed
located at the same redshift as the extended \lya emission and that it
is a massive galaxy ($M_{\rm dyn}$ = 3--7\E{11}\msun) based on the broad
CO line-width (FWHM = 700\,\kms).

Combined with Keck/LRIS longslit spectroscopy of \lya and \heii, we
constrain the kinematics of the extended gas using the CO emission as the
best tracer of the systemic redshift.  At the position of the MIPS source,
the \lya profile is symmetric and its line center agrees with those of
the CO lines.  This implies that there are no significant bulk flows
and the photo-ionization from the MIPS source might be the dominant
source of the \lya emission.  Near the peak of the \lya nebula, the
gas is slowly receding ($\sim$100\kms) with respect to the MIPS source,
thus disfavoring the hyper-/superwind hypothesis where extreme galactic
winds are responsible for the extended \lya emission.  However,
we note that we cannot rule out the possibility that gas flow within
LABd05 is tangential to our LOS.

We find that a significantly sub-thermal line ratio between
the two CO transitions, \ICO{5}{4}/\ICO{3}{2} = 0.97 $\pm$ 0.21
[\LpCO{5}{4}/\LpCO{3}{2} = 0.35 $\pm$ 0.08]. This line ratio is lower than
the average values found in high-$z$ SMGs and QSOs, but is consistent
with the value found in the center of the Milky Way.  This line ratio
indicates that there is a large reservoir of low-density molecular gas
that could be spread over the vicinity of the MIPS source.  Observations
of CO(7--6) and CO(1--0) lines with higher spatial resolution are required
to further constrain the properties of the star-forming regions within
this \lya blob.

\medskip
\acknowledgments

We thank Chin Shin Chang, Tessel van der Laan and Jan Martin Winters for
supporting our PdBI observations. YY thank Alexander Karim and Benjamin
Magnelli for helpful discussions.  We thank the anonymous referee
for a thorough reading of the manuscript and for providing helpful
suggestions.
This work is based on observations carried out with the IRAM Plateau
de Bure Interferometer and the IRAM-30m Telescope. IRAM is supported by
INSU/CNRS (France), MPG (Germany) and IGN (Spain).
This work was carried out within the Collaborative Research Council 956,
sub-project A1, funded by the Deutsche Forschungsgemeinschaft (DFG).
Support for RD was provided by the DFG priority program 1573, ``The
Physics of the Interstellar Medium''.
M.K.M.P.~was supported by a Dark Cosmology Centre Fellowship.  
AD's research activities are supported by NOAO, which is operated by the
Association of Universities for Research in Astronomy under a cooperative
agreement with the US National Science Foundation.  AD's research is
also supported in part by the Radcliffe Institute for Advanced Study at
Harvard University.
Some of the data presented herein were obtained at the W.M. Keck
Observatory, which is operated as a scientific partnership among the
California Institute of Technology, the University of California and the
National Aeronautics and Space Administration. The Observatory was made
possible by the generous financial support of the W.M. Keck Foundation.

\bigskip
Facilities: \facility{PdBI, Keck(LRIS)}

\def\arraystretch{1.0}
\newcommand\ff[1]{\tablenotemark{#1}}
\begin{deluxetable}{lccccccc cc}
\tablewidth{0pt}
\tabletypesize{\small}
\tablecaption{CO Line Observations}
\tablehead{
\colhead{Source          }&
\colhead{Transition      }&
\colhead{$\nu_{\rm obs}$ }&    
\colhead{$S_{\rm cont}$  }&
\colhead{$z_{\rm CO}$    }&
\colhead{FWHM            }&
\colhead{$S \Delta V$    }&
\colhead{\lco            }&
\colhead{\lpco           }\\[0.5ex]
\colhead{                }&
\colhead{                }&
\colhead{(GHz)           }&
\colhead{(mJy)           }&
\colhead{                }&
\colhead{(\kms)          }&
\colhead{(Jy\,\kms)      }&
\colhead{($10^{7}$ \lsun)}&
\colhead{($10^{10}$\unitlpco)}
}
\startdata
LABd05              &  5--4  &  157.6215 &       0.38 $\pm$      0.05   &     2.6560 $\pm$    0.0006 &      734 $\pm$    124 &     1.18 $\pm$     0.17        &     9.37  &      1.53 \\
                    &  3--2  &\bb94.5831 &            $ < $      0.15   &     2.6560 $\pm$    0.0007 &      676 $\pm$    129 &     1.22 $\pm$     0.20        &     5.78  &      4.36 \\
\hline\\[-1.5ex]                                                                                                                                                        
SSA22--LAB18\ff{a}  &  3--2  &\bb84.5500 &            $ < $      0.13   &     \nodata                &      \nodata          &          $ < $     0.27\ff{b}  &  $<$1.67 &  $<$  1.26 
\enddata
\label{tab:CO}
\tablenotetext{a}{Upper limits are 3\sig.}
\tablenotetext{b}{FWHM = 400\,\kms assumed.\vspace{0.5cm}}
\end{deluxetable}

\appendix
\section{Observations of SSA22-LAB18}

\begin{figure*}
\vspace{-0.2cm}
\epsscale{0.9}
\plotone{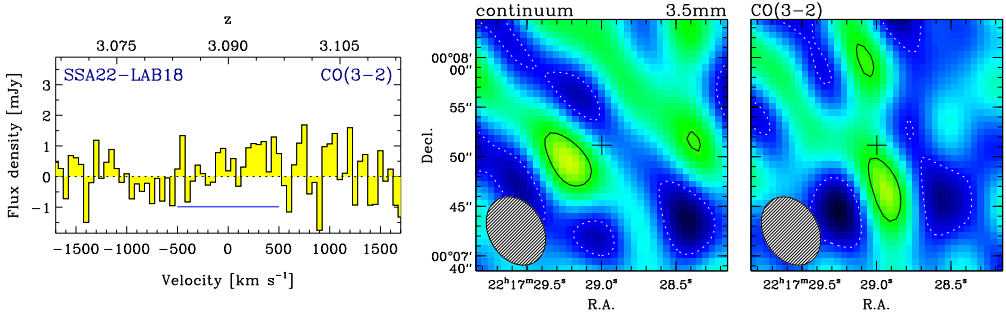}
\caption{
PdBI CO line observations for SSA22--LAB18. Neither the dust continuum
nor the CO(3--2) line is detected.
({\it Left})
Spectra at the position and redshift of the expected CO emission with a
velocity resolution of 50\,\kms. The velocity is relative to the tuned
frequency of the receiver corresponding to the central wavelength of
the narrowband filter used in its discovery \citep{Matsuda04}.
({\it Middle})
Dirty continuum image. 
({\it Right})
Line image integrated over 1000\kms. See the caption of Fig.~\ref{fig:CO}
for details.
}
\label{fig:CO_LAB18}
\end{figure*}

\begin{figure}
\epsscale{0.65}
\plotone{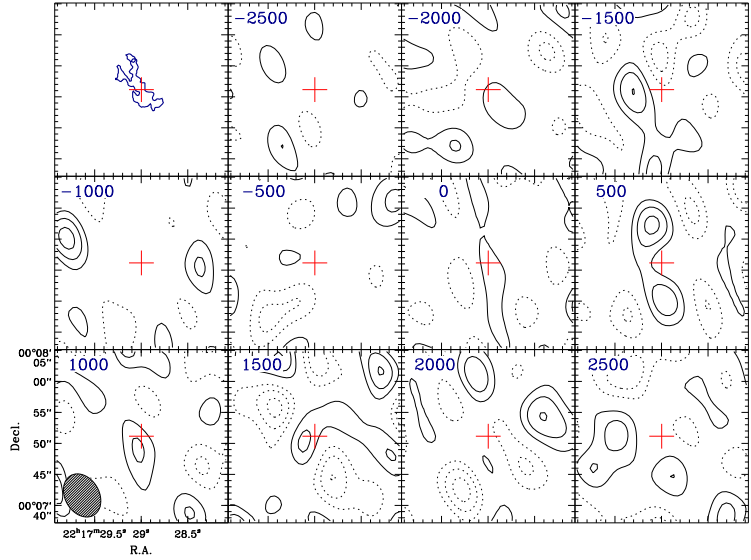} 
\caption{
Integrated channel map of SSA22-LAB18. Each channel is integrated over a
500\,\kms\ bandwidth centered at the velocities shown in the upper left
corners.  The contours are at $-$3, $-$2, $-$1, 1, 2, and 3$\sigma$,
where $\sigma$ is the rms noise per 500\kms\ velocity bin (0.21 mJy
beam$^{-1}$).  The phase center is marked with a cross in each panel. The
first panel in the upper left corner shows a \lya contour corresponding
to a surface brightness of $\sim$2.2\E{-18}\,\unitcgssb \citep{Matsuda04}.
} 
\label{fig:map_LAB18}
\end{figure}

Here we summarize the results of the observations of SSA22-LAB18.
This \lya blob was identified from a narrowband imaging survey at $z=3.1$
\citep{Matsuda04}.  At the time of our observation, this source was known
to have the brightest FIR luminosity among \lya blobs \citep{Geach05,
Yang12}.  Figure \ref{fig:CO_LAB18} shows the PdBI CO line observations
for SSA22--LAB18. For reference, we also display the entire dataset
in Figure \ref{fig:map_LAB18}.  Because the spectroscopic redshift of
this \lya blob is not available, we searched for detections over the
entire redshift range ($z$ = 3.06 -- 3.13) covered by the narrowband
filter ({\sl NB}497).  Neither dust continuum or CO(3--2) line emission
is detected within the \lya blob.

\begin{figure}
\epsscale{0.65}
\vspace{-0.1cm}
\plotone{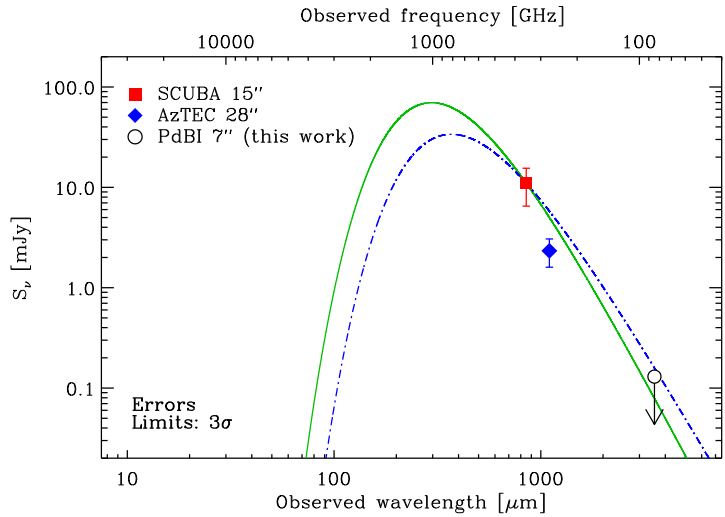} 
\caption{
FIR flux densities of SSA22-LAB18. Different symbols represent the (sub)mm
observations with different instruments and beams sizes \cite[][this
work]{Geach05, Tamura13}.  All the uncertainties and upper limits are shown
at the 3$\sigma$ level.  The solid and dot-dashed lines represent the modified
blackbody SEDs normalized at $S_{850\um}$ = 11\,mJy with ($T_d$,
$\beta$) = (40\,K, 2) and (30\,K, 1.5), respectively.
} 
\label{fig:SED_LAB18}
\end{figure}

We compare the non-detection in 3.5mm continuum with the existing FIR flux
measurements. Figure \ref{fig:SED_LAB18} show the FIR flux densities from
SCUBA \citep{Geach05}, AzTEC-ASTE \citep{Tamura13} and our measurements.
Different symbols represent the (sub)mm observations with different
instruments and beams sizes. For the AzTEC-ASTE measurement, we show the
measurement extracted at the position of SSA22--LAB18--b, one of the two
components in SSA22--LAB18 \citep{Tamura13}.  Note that SSA22--LAB18--ab
are blended with a bright nearby source, thus the actual flux might be
even lower than reported.  For the illustration purposes, the solid and
dot-dashed lines represent the modified blackbody SEDs normalized at
the SCUBA photometry, $S_{850\um}$ = 11\,mJy with ($T_d$, $\beta$) =
(40\,K, 2) and (30\,K, 1.5), respectively.  Our upper limit at 3.5mm
is not stringent enough to determine which of the previous continuum
measurements is correct.

\begin{thebibliography}{}
 \bibitem[Alloin et al.(2007)]{Alloin07} Alloin, D., Kneib, J.-P., Guilloteau, S., \& Bremer, M.\ 2007, \aap, 470, 53 
 \bibitem[Barvainis et al.(1997)]{Barvainis97} Barvainis, R., Maloney, P., Antonucci, R., \& Alloin, D.\ 1997, \apj, 484, 695
 \bibitem[Beelen et al.(2006)]{Beelen06} Beelen, A., Cox, P., Benford, D.~J., et al.\ 2006, \apj, 642, 694 
 \bibitem[Bower et al.(2004)]{Bower04} Bower, R.~G., et al.\ 2004, \mnras, 351, 63 
 \bibitem[Bridge et al.(2013)]{Bridge13} Bridge, C.~R., Blain, A., Borys, C.~J.~K., et al.\ 2013, \apj, 769, 91 
 \bibitem[Bussmann et al.(2009)]{Bussmann09} Bussmann, R.~S., et al.\ 2009, \apj, 705, 184 
 \bibitem[Carilli \& Walter(2013)]{Carilli&Walter13} Carilli, C.~L., \& Walter, F.\ 2013, \araa, 51, 105 
 \bibitem[Chapman et al.(2004)]{Chapman04} Chapman, S.~C., Scott, D., Windhorst, R.~A., Frayer, D.~T., Borys, C., Lewis, G.~F., \& Ivison, R.~J.\ 2004, \apj, 606, 85 
 \bibitem[Colbert et al.(2011)]{Colbert11} Colbert, J.~W., Scarlata, C., Teplitz, H., Francis, P., Palunas, P., Williger, G.~M., \& Woodgate, B.\ 2011, \apj, 728, 59 
 \bibitem[Danielson et al.(2011)]{Danielson11} Danielson, A.~L.~R., Swinbank, A.~M., Smail, I., et al.\ 2011, \mnras, 410, 1687 
 \bibitem[Dannerbauer et al.(2009)]{Dannerbauer09} Dannerbauer, H., Daddi, E., Riechers, D.~A., et al.\ 2009, \apjl, 698, L178 
 \bibitem[Dey et al.(2005)]{Dey05} Dey, A., et al.\ 2005, \apj, 629, 654 
 \bibitem[Dey et al.(2008)]{Dey08} Dey, A., Soifer, B.~T., Desai, V., et al.\ 2008, \apj, 677, 943 
 \bibitem[Dijkstra et al.(2006a)]{Dijkstra06a} Dijkstra, M., Haiman, Z., \& Spaans, M.\ 2006, \apj, 649, 14 
 \bibitem[Dijkstra et al.(2006b)]{Dijkstra06b} Dijkstra, M., Haiman, Z., \& Spaans, M.\ 2006, \apj, 649, 37 
 \bibitem[Dijkstra \& Loeb(2009)]{Dijkstra&Loeb09} Dijkstra, M., \& Loeb, A.\ 2009, \mnras, 400, 1109 
 \bibitem[Erb et al.(2011)]{Erb11} Erb, D.~K., Bogosavljevi{\'c}, M., \& Steidel, C.~C.\ 2011, \apjl, 740, L31 
 \bibitem[Fardal et al.(2001)]{Fardal01} Fardal, M.~A., Katz, N., Gardner, J.~P., Hernquist, L., Weinberg, D.~H., \& Dav{\' e}, R.\ 2001, \apj, 562, 605 
 \bibitem[Farina et al.(2013)]{Farina13} Farina, E.~P., Falomo, R., Decarli, R., Treves, A., \& Kotilainen, J.~K.\ 2013, \mnras, 429, 1267 
 \bibitem[Fixsen et al.(1999)]{Fixsen99} Fixsen, D.~J., Bennett, C.~L., \& Mather, J.~C.\ 1999, \apj, 526, 207 
 \bibitem[Francis et al.(2001)]{Francis01} Francis, P.~J., et al.\ 2001, \apj, 554, 1001 
 \bibitem[Frayer et al.(1999)]{Frayer99} Frayer, D.~T., Ivison, R.~J., Scoville, N.~Z., et al.\ 1999, \apjl, 514, L13 
 \bibitem[Geach et al.(2005)]{Geach05} Geach, J.~E., et al.\ 2005, \mnras, 363, 1398
 \bibitem[Geach et al.(2009)]{Geach09} Geach, J.~E., et al.\ 2009, \apj, 700, 1 
 \bibitem[Goerdt et al.(2010)]{Goerdt10} Goerdt, T., Dekel, A., Sternberg, A., Ceverino, D., Teyssier, R., \& Primack, J.~R.\ 2010, \mnras, 407, 613 
 \bibitem[Greve et al.(2005)]{Greve05} Greve, T.~R., et al.\ 2005, \mnras, 359, 1165 
 \bibitem[Hainline et al.(2004)]{Hainline04} Hainline, L.~J., Scoville, N.~Z., Yun, M.~S., et al.\ 2004, \apj, 609, 61 
 \bibitem[Haiman, Spaans, \& Quataert(2000)]{Haiman00} Haiman, Z., Spaans, M., \& Quataert, E.\ 2000, \apjl, 537, L5 
 \bibitem[Hashimoto et al.(2013)]{Hashimoto13} Hashimoto, T., Ouchi, M., Shimasaku, K., et al.\ 2013, \apj, 765, 70 
 \bibitem[Hayes et al.(2011)]{Hayes11} Hayes, M., Scarlata, C., \& Siana, B.\ 2011, \nat, 476, 304 
 \bibitem[Hennawi \& Prochaska(2013)]{Hennawi13} Hennawi, J.~F., \& Prochaska, J.~X.\ 2013, \apj, 766, 58 
 \bibitem[Ivison et al.(1998)]{Ivison98} Ivison, R.~J., Smail, I., Le Borgne, J.-F., et al.\ 1998, \mnras, 298, 583 
 \bibitem[Ivison et al.(2011)]{Ivison11} Ivison, R.~J., Papadopoulos, P.~P., Smail, I., et al.\ 2011, \mnras, 412, 1913 
 \bibitem[Keel et al.(1999)]{Keel99} Keel, W.~C., Cohen, S.~H., Windhorst, R.~A., \& Waddington, I.\ 1999, \aj, 118, 2547  
 \bibitem[Kohno et al.(2008)]{Kohno08} Kohno, K., et al.\ 2008, Panoramic Views of Galaxy Formation and Evolution, 399, 264 
 \bibitem[Kulas et al.(2012)]{Kulas12} Kulas, K.~R., Shapley, A.~E., Kollmeier, J.~A., et al.\ 2012, \apj, 745, 33 
 \bibitem[Leroy et al.(2011)]{Leroy11} Leroy, A.~K., Bolatto, A., Gordon, K., et al.\ 2011, \apj, 737, 12 
 \bibitem[Neri et al.(2003)]{Neri03} Neri, R., Genzel, R., Ivison, R.~J., et al.\ 2003, \apjl, 597, L113 
 \bibitem[Matsuda et al.(2004)]{Matsuda04} Matsuda, Y., et al.\ 2004, \aj, 128, 569 
 \bibitem[Matsuda et al.(2007)]{Matsuda07} Matsuda, Y., Iono, D., Ohta, K., Yamada, T., Kawabe, R., Hayashino, T., Peck, A.~B., \& Petitpas, G.~R.\ 2007, \apj, 667, 667 
 \bibitem[Matsuda et al.(2011)]{Matsuda11} Matsuda, Y., et al.\ 2011, \mnras, 410, L13      
 \bibitem[McLinden et al.(2011)]{McLinden11} McLinden, E.~M., Finkelstein, S.~L., Rhoads, J.~E., et al.\ 2011, \apj, 730, 136 
 \bibitem[McLinden et al.(2013)]{McLinden13} McLinden, E.~M., Malhotra, S., Rhoads, J.~E., et al.\ 2013, \apj, 767, 48 
 \bibitem[Melbourne et al.(2012)]{Melbourne12} Melbourne, J., Soifer, B.~T., Desai, V., et al.\ 2012, \aj, 143, 125 
 \bibitem[Ouchi et al.(2009)]{Ouchi09} Ouchi, M., et al.\ 2009, \apj, 696, 1164
 \bibitem[Overzier et al.(2013)]{Overzier13} Overzier, R.~A., Nesvadba, N.~P.~H., Dijkstra, M., et al.\ 2013, arXiv:1305.2926 
 \bibitem[Prescott et al.(2008)]{Prescott08} Prescott, M.~K.~M., Kashikawa, N., Dey, A., \& Matsuda, Y.\ 2008, \apjl, 678, L77 
 \bibitem[Prescott et al.(2009)]{Prescott09} Prescott, M.~K.~M., Dey, A., \& Jannuzi, B.~T.\ 2009, \apj, 702, 554 
 \bibitem[Prescott et al.(2012a)]{Prescott12a} Prescott, M.~K.~M., Dey, A., \& Jannuzi, B.~T.\ 2012, \apj, 748, 125 
 \bibitem[Prescott et al.(2012b)]{Prescott12b} Prescott, M.~K.~M., Dey, A., Brodwin, M., et al.\ 2012, \apj, 752, 86 
 \bibitem[Riechers et al.(2011a)]{Riechers11a} Riechers, D.~A., Cooray, A., Omont, A., et al.\ 2011, \apjl, 733, L12
 \bibitem[Riechers et al.(2011b)]{Riechers11b} Riechers, D.~A., Hodge, J., Walter, F., Carilli, C.~L., \& Bertoldi, F.\ 2011, \apjl, 739, L31 
 \bibitem[Saito et al.(2006)]{Saito06} Saito, T., Shimasaku, K., Okamura, S., Ouchi, M., Akiyama, M., \& Yoshida, M.\ 2006, \apj, 648, 54 
 \bibitem[Schumacher et al.(2012)]{Schumacher12} Schumacher, H., Mart{\'{\i}}nez-Sansigre, A., Lacy, M., Rawlings, S., \& Schinnerer, E.\ 2012, \mnras, 423, 2132
 \bibitem[Smith \& Jarvis(2007)]{Smith&Jarvis07} Smith, D.~J.~B., \& Jarvis, M.~J.\ 2007, \mnras, 378, L49 
 \bibitem[Solomon et al.(1997)]{Solomon97} Solomon, P.~M., Downes, D., Radford, S.~J.~E., \& Barrett, J.~W.\ 1997, \apj, 478, 144 
 \bibitem[Steidel et al.(2000)]{Steidel00} Steidel, C.~C., Adelberger, K.~L., Shapley, A.~E., Pettini, M., Dickinson, M., \& Giavalisco, M.\ 2000, \apj, 532, 170 
 \bibitem[Steidel et al.(2004)]{Steidel04} Steidel, C.~C., Shapley, A.~E., Pettini, M., Adelberger, K.~L., Erb, D.~K., Reddy, N.~A., \& Hunt, M.~P.\ 2004, \apj, 604, 534 
 \bibitem[Steidel et al.(2010)]{Steidel10} Steidel, C.~C., Erb, D.~K., Shapley, A.~E., Pettini, M., Reddy, N., Bogosavljevi{\'c}, M., Rudie, G.~C., \& Rakic, O.\ 2010, \apj, 717, 289 
 \bibitem[Steidel et al.(2011)]{Steidel11} Steidel, C.~C., Bogosavljevi{\'c}, M., Shapley, A.~E., et al.\ 2011, \apj, 736, 160 
 \bibitem[Swinbank et al.(2010)]{Swinbank10} Swinbank, A.~M., Smail, I., Longmore, S., et al.\ 2010, \nat, 464, 733 
 \bibitem[Tacconi et al.(2006)]{Tacconi06} Tacconi, L.~J., Neri, R., Chapman, S.~C., et al.\ 2006, \apj, 640, 228
 \bibitem[Tamura et al.(2013)]{Tamura13} Tamura, Y., Matsuda, Y., Ikarashi, S., et al.\ 2013, \mnras, 722 
 \bibitem[Taniguchi \& Shioya(2000)]{Taniguchi&Shioya00} Taniguchi, Y.~\& Shioya, Y.\ 2000, \apjl, 532, L13 
 \bibitem[Verhamme et al.(2006)]{Verhamme06} Verhamme, A., Schaerer, D., \& Maselli, A.\ 2006, \aap, 460, 397 
 \bibitem[Verhamme et al.(2008)]{Verhamme08} Verhamme, A., Schaerer, D., Atek, H., \& Tapken, C.\ 2008, \aap, 491, 89         
 \bibitem[Wagg et al.(2012)]{Wagg12b} Wagg, J., Pope, A., Alberts, S., et al.\ 2012, \apj, 752, 91
 \bibitem[Wagg \& Kanekar(2012)]{Wagg12a} Wagg, J., \& Kanekar, N.\ 2012, \apjl, 751, L24 
 \bibitem[Weijmans et al.(2009)]{Weijmans09} Weijmans, A.-M., Bower, R.~G., Geach, J.~E., Swinbank, A.~M., Wilman, R.~J., de Zeeuw, P.~T., \& Morris, S.~L.\ 2009, \mnras, 1911
 \bibitem[Wei{\ss} et al.(2003)]{Weiss03} Wei{\ss}, A., Henkel, C., Downes, D., \& Walter, F.\ 2003, \aap, 409, L41
 \bibitem[Wei{\ss} et al.(2005)]{Weiss05} Wei{\ss}, A., Walter, F., \& Scoville, N.~Z.\ 2005, \aap, 438, 533 
 \bibitem[Wei{\ss} et al.(2007)]{Weiss07} Wei{\ss}, A., Downes, D., Walter, F., \& Henkel, C.\ 2007, From Z-Machines to ALMA: (Sub)Millimeter Spectroscopy of Galaxies, 375, 25 
 \bibitem[Wei{\ss} et al.(2009)]{Weiss09} Wei{\ss}, A., Ivison, R.~J., Downes, D., et al.\ 2009, \apjl, 705, L45
 \bibitem[Wilman et al.(2005)]{Wilman05} Wilman, R.~J., Gerssen, J., Bower, R.~G., Morris, S.~L., Bacon, R., de Zeeuw, P.~T., \& Davies, R.~L.\ 2005, \nat, 436, 227 
 \bibitem[Yamada et al.(2012)]{Yamada12} Yamada, T., Matsuda, Y., Kousai, K., et al.\ 2012, arXiv:1203.3633 
 \bibitem[Yang et al.(2009)]{Yang09} Yang, Y., Zabludoff, A., Tremonti, C., Eisenstein, D., \& Dav{\'e}, R.\ 2009, \apj, 693, 1579 
 \bibitem[Yang et al.(2010)]{Yang10} Yang, Y., Zabludoff, A., Eisenstein, D., \& Dav{\'e}, R.\ 2010, \apj, 719, 1654 
 \bibitem[Yang et al.(2011)]{Yang11} Yang, Y., Zabludoff, A., Jahnke, K., Eisenstein, D., Dav{\'e}, R., Shectman, S.~A., \& Kelson, D.~D.\ 2011, \apj, 735, 87 
 \bibitem[Yang et al.(2012)]{Yang12} Yang, Y., Decarli, R., Dannerbauer, H., et al.\ 2012, \apj, 744, 178 
\end{thebibliography}
\end{document}